\documentclass[conference,10pt]{IEEEtran}
\usepackage{graphicx,color}
\usepackage{amsmath, amsthm, amsfonts, amssymb, amsbsy,nccmath,bbm}
\usepackage{algorithm}
\usepackage{algpseudocode}
\usepackage{enumerate}
\usepackage{lipsum}
\newtheorem{lemma}{Lemma}

\usepackage[sort,compress]{cite}
\usepackage{epsfig}
\usepackage{epstopdf}
\usepackage{mathtools}
\usepackage{dsfont}
\usepackage{epstopdf}

\usepackage[inline]{enumitem}   
\makeatletter
\newcommand{\inlineitem}[1][]{%
\ifnum\enit@type=\tw@
    {\descriptionlabel{#1}}
  \hspace{\labelsep}%
\else
  \ifnum\enit@type=\z@
       \refstepcounter{\@listctr}\fi
    \quad\@itemlabel\hspace{\labelsep}%
\fi}
\makeatother
\newcommand{\beq}{\begin{equation}}
\newcommand{\eeq}{\end{equation}}

\def\adots{\mathinner{\mskip0mu\raise0pt\vbox{\kern7pt\hbox{.}}\mskip3mu
          \raise4pt\hbox{.}\mskip3mu\raise8pt\hbox{.}\mskip0mu}}

\newcommand{\bmW}{{\mathbf W}}
\newcommand{\bmV}{{\mathbf V}}

\newcommand{\tr}{\mbox{tr}}

\newcommand{\bmh}{\bfh}

\usepackage{bm}

\newcommand{\bmtheta}{{\bm \theta}}

\newcommand{\bmx}{{\bm x}}
\newcommand{\bmy}{{\bm y}}
\newcommand{\bmv}{{\bm v}}

\newcommand{\bmF}{{\bm F}}

\newcommand{\bmH}{{\bm H}}
\newcommand{\bmQ}{{\bm Q}}

\renewcommand{\bmh}{{\bm h}}

\newcommand{\bmA}{{\bm A}}

\newcommand{\bmC}{{\bf C}}

\newcommand{\bmHh}{\widehat{\bmH}}
\newcommand{\bmhh}{\widehat{\bmh}}
\newcommand{\bmI}{{\bf {I}}}
\newcommand{\bmD}{{\bf {D}}}
\newcommand{\bmY}{{\bf {Y}}}

\newcommand{\bmm}{{\bf {m}}}
\newcommand{\bmzero}{{\bm 0}}

\makeatletter
\newcommand\fs@spaceruled{\def\@fs@cfont{\bfseries}\let\@fs@capt\floatc@ruled
  \def\@fs@pre{\vspace{0.5\baselineskip}\hrule height.8pt depth0pt \kern2pt}%
  \def\@fs@post{\kern1pt\hrule\relax}%
  \def\@fs@mid{\kern2pt\hrule\kern2pt}%
  \let\@fs@iftopcapt\iftrue}
\makeatother

\newcommand{\bit}{\begin{itemize}}
\newcommand{\eit}{\end{itemize}}

\newcommand{\bmz}{\mathbf{z}}

\renewcommand{\bmh}{{\mathbf h}}
\renewcommand{\bmh}{{\mathbf h}}

\newcommand{\kbar}{\overline{k}}

\newcommand{\ibar}{\overline{i}}

\renewcommand{\bmA}{{\mathbf A}}

\newcommand{\bmX}{{\mathbf X}}

\newcommand{\bmB}{{\mathbf B}}

\newcommand{\lpbar}{{\overline{l'}}}

\newcommand{\bmPsi}{{\bm \Psi}}

\newcommand{\bmLambda}{{\boldsymbol{\Lambda}}}

\usepackage{lipsum}

\newcommand{\bmXi}{{\boldsymbol \Xi}}

\renewcommand{\bmzero}{{\boldsymbol 0}}
\newcommand{\bmone}{{\boldsymbol 1}}

\DeclareMathOperator{\E}{\mathbb{E}}

\usepackage{subfiles}

\usepackage[acronym]{glossaries}

\newacronym{kld}{KLD}{Kullback–Leibler divergence}
\newacronym{snr}{SNR}{signal-to-noise ratio}
\newacronym{ap}{AP}{access point}

\newtheorem{property}{Property}
\newtheorem{assumption}{Assumption}
\begin{document}
\linespread{0.82}

\title{Decentralized Expectation Propagation for Semi-\\ Blind Channel Estimation in Cell-Free Networks}
\author{%
  \IEEEauthorblockN{Zilu Zhao,  Dirk Slock}
  \IEEEauthorblockA{
			\small
			Communication Systems Department, EURECOM, France \\	
			zilu.zhao@eurecom.fr, dirk.slock@eurecom.fr
			\vspace{-3mm}		}
	}

\maketitle

\begin{abstract}
This paper serves as a correction to the conference version \cite{EURECOM+7816}. In this work, we explore uplink communication in cell-free (CF) massive multiple-input multiple-output (MaMIMO) systems, employing semi-blind transmission structures to mitigate pilot contamination. We propose a simplified, decentralized method based on Expectation Propagation (EP) for semi-blind channel estimation. By utilizing orthogonal pilots, we preprocess the received signals to establish a simplified equivalent factorization scheme for the transmission process.
Moreover, this study integrates Central Limit Theory (CLT) with EP, eliminating the need to introduce new auxiliary variables in the factorization scheme. We also refine the algorithm by assessing the variable scales involved. Finally, a decentralized approach is proposed to significantly reduce the computational demands on the Central Processing Unit (CPU).
\end{abstract}
\vspace{-1mm}
\section{Introduction}
\label{Intro}
\vspace{-1mm}


One of the unique features of Cell-Free (CF) Massive MIMO (MaMIMO) networks is user terminals (UTs) in a given area are served by all the access points (APs) in the same area. This leads to the problem of pilot contamination. Addressing this issue, Semi-Blind approaches \cite{gholami2021tackling} have been explored to mitigate the effects of pilot contamination. In \cite{gholami2021tackling}, the authors consider a deterministic approach treating the channel coefficients as deterministic unknown parameters and delve into the analysis of the Cramer-Rao bound and identifiability.
 
In the context of Bayesian inference, the Semi-Blind approach is modeled as a bilinear inference problem, where the APs must jointly estimate both the channel coefficients and the user signals. Message-passing algorithms play a critical role here. (Loopy) Belief Propagation (BP) \cite{wainwright2008graphical} is a widely used method in Bayesian inference. It is an iterative method exploiting the structure for a given factorization scheme (e.g., joint/posterior distribution) to lower the computational loads. One of the most potent message-passing algorithms is Expectation Propagation (EP) \cite{minka2001family}. Besides exploiting the factorization scheme, EP also projects the beliefs (marginal posterior) to a family of simple distributions. 
We can consider BP a special case of EP, in which we assume the projection destination set to be the set of all distributions.

A centralized iterative algorithm has been explored based on variable level EP (VL-EP) \cite{gholami2021message} in which the authors assume Gaussian input and combine VL-EP with Expectation Maximization (EM). A more recent development is the distributed method proposed in \cite{karataev2024bilinear}. This approach distributes the computational load at the Central Process Unit (CPU) by enabling each Access Point (AP) to carry out part of the computation.

In \cite{takeuchi2023decentralized}, the author introduces Decentralized Generalized Approximate Message Passing (D-GAMP). This method is a hybrid of Consensus Propagation \cite{moallemi2006consensus} and Approximate Message Passing (AMP), effectively eliminating the need for CPU.

\subsection{Main Contributions}

This paper presents a simplified, decentralized, EP-based method designed to address the Semi-Blind estimation problem in communication systems. By utilizing orthogonal pilots, we are able to decouple the channels for different users into mutually exclusive groups, which reduces computational complexity. To further decrease computational demands, we integrate Expectation Propagation (EP) with Central Limit Theory (CLT), treating the interference as noise. Drawing inspiration from \cite{parker2014bilinear}, we introduce further simplifications through scale analysis. Additionally, to lessen the load on the central processing unit (CPU), we explore a decentralized scheme.

\section{System Model}

We consider a semi-blind signal model containing $L$ APs. At the $l$-th AP,\vspace{-2mm}
\beq
\vspace{-2mm}
    \begin{bmatrix}
        \bmY_{p, l} &\bmY_{l}
    \end{bmatrix}
    =
    \bmH_{l}
    \begin{bmatrix}
        \bmX_{p} &\bmX
    \end{bmatrix}
    +
    \begin{bmatrix}
        \bmV_{p, l} & \bmV_{l}
    \end{bmatrix}
    .
\eeq
The received signals are composed of pilot part $\bmY_{p, l} \in \mathbb{C}^{N \times P}$ and data part $\bmY_{l}\in \mathbb{C}^{N \times T}$. The channels between different users are considered independent Gaussian i.e. $\text{vec}{(\bmH_l)}\sim \mathcal{CN}(\bmzero, \bmXi_{l})$ where $\bmXi_{l} \in \mathbb{C}^{NK\times NK}$ is a block diagonal matrix of $K$ blocks $\bmXi_{\bmh_{lk}}\in \mathbb{C}^{N\times N}$. The transmitted symbols can be decomposed as pilot symbols $\bmX_{p}\in \mathcal{S}^{K\times P}$ and data symbols $\bmX\in \mathcal{S}^{K\times T}$, where $\mathcal{S}$ is the constellation set. We assume that the elements $x_{kt}$ in $\bmX$ follow the categorical distribution $p(x_{kt})$. The signal power is denoted as $\sigma_x^2$. The noise is considered as i.i.d. Gaussian distribution, and thus, $\text{vec}(\begin{bmatrix}
        \bmV_{p, l} & \bmV_{l}
    \end{bmatrix})\sim\mathcal{CN}(0, \sigma_{v}^2\bmI)$.


\subsection {Orthogonal Pilot sequences}
If orthogonal pilot sequences are used, 
we can first preprocess the pilot observation by right multiplying it with $\bmx_{p, g}^*$ which is the conjugated $g$-th pilot sequence. This results in an equivalent observation $\bmy_{p, lg}$
\vspace{-2mm}
\beq
\vspace{-2mm}
    \bmy_{p, lg}=\bmY_{p, l}\bmx_{p, g}^*=\sum_{k\in G_g}P \sigma_x^2\bmh_{lk}+\bmv_{p, lg}
\eeq
where $\bmv_{p, lg}=\bmV_{p, l}\bmx_{p, g}^*\sim \mathcal{N}(\bmv_{p, lg}|\bmzero, P\sigma_x^2\sigma_v^2\bmI)$, $G_g$ denote the set of users using the $g$-th pilot sequence. We observe that every $\bmh_{lk}$ occurs only in one group $G_g$, and the cross-correlation $\E[\bmv_{p,lg}\bmv_{p, lg'}^H]$ is an all-zero matrix for all $g\neq g'$. Therefore, the observations $\bmy_{p, lg}$ and $\bmy_{p, lg'}$ are independent. 
With orthogonal pilots, the factorization scheme is derived as
\begin{align}
        &p(\{\bmy_{p, lg}\}, \{\bmY_{l}\}, \{\bmH_{l}\}, \bmX,  \{\bmV_{l}\}) \label{eq:isit243}\\
        =&\prod_{k,t} p(x_{kt})\prod_{l}\prod_{t_{1}=1}^T p(\bmy_{lt_{1}}|\bmH_{l}, \bmx_{:t_{1}})
        \prod_{g} p(\bmy_{p,lg},\bmH_{lg}) \nonumber        
\end{align}
where $\bmH_{lg}$ is a matrix collecting all $k\in G_{g}, \bmh_{lk}$ as its column vectors, and $\bmx_{:t_1}$ denotes the $t_1$-th column of $\bmX$. We will base our EP (BP) algorithm based on this factorization scheme.

\section{Expectation Propagation Overview}
EP approximates the factors in a factorization scheme to simpler ones \cite{ngo2020multi}.  With a given factorization, the update algorithm in EP can be interpreted as message passing of two types of messages, i.e., the message $\mu_{\Psi; \theta_i}(\theta_i)$ from factor node $\Psi$ to variable node $\theta_i$ and the message $\mu_{\theta_i;\Psi}(\theta_i)$ from variable $\theta_i$ to factor $\Psi$: \cite{zou2018concise}
\beq
\begin{split}
    \mu_{\theta_i;\Psi}(\theta_i) \propto \prod_{\Phi \neq \Psi}\mu_{\Phi;\theta_i}(\theta_i);\;
    \mu_{\Psi; \theta_i}(\theta_i) \propto \frac{\text{proj}(b_{\Psi}(\theta_i))}{\mu_{\theta_i; \Psi}(\theta_i)},
\end{split}
\label{eq:ISIT246}
\eeq
where $b_{\Psi}(\theta_i)$ is the belief of $\theta_i$ at node $\Psi$:
\vspace{-1mm}
\beq
\vspace{-1mm}
    b_{\Psi}(\theta_i)\propto\mu_{\theta_i; \Psi}(\theta_i)\int \Psi(\bmtheta) \prod_{j\neq i} \mu_{\theta_j; \Psi}(\theta_j) d\bmtheta_{\ibar}.
    \label{eq:EPexplainationBelief}
\eeq
The notation $\bmtheta_{\ibar}$ denotes all elements in $\bmtheta$ except the $i$-th one.
The operation $\text{proj}(p)$ project a given distribution $p$ into a target family $Q$ \cite{zou2018concise}, i.e.,
\vspace{-1mm}
\beq
\vspace{-1mm}
    \text{proj}(p)=\arg\min_{q\in Q}KLD(p\| q),
\eeq
where $KLD(p\|q)=\int p(\theta) \ln \frac{p(\theta)}{q(\theta)} d\theta$ is the Kullback–Leibler divergence.

BP, on the other hand, can be considered a special form of EP. The only difference between BP and EP is that there is no projection step in BP. We assume all the messages in this paper are normalized to $1$.

\subsection {Expectation Propagation on Semi-Blind structure}
\begin{figure}[t]
    \centering
    \includegraphics[width=0.48\textwidth]{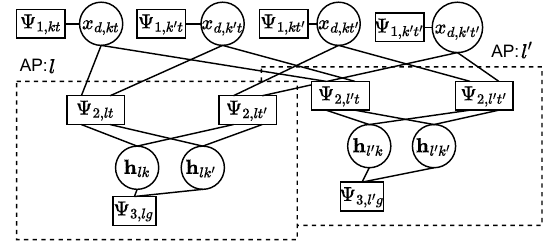}
    \caption{Partial factor graph}
    \label{fig:FG}
    \vspace{-6mm}
\end{figure}

For simplicity, we denote the factors in the factorization scheme \eqref{eq:isit243} as
    \begin{align}
        \Psi_{1, kt}\!=p(x_{kt});\;
        \Psi_{2, lt}\!=p(\bmy_{lt}|\bmH_{l}, x_{:t});\;\Psi_{3, lg}\!\!=p(\bmy_{p, lg},\bmH_{lg}).\nonumber
    \end{align}
    The factor graph for \eqref{eq:isit243} is illustrated in Fig. \ref{fig:FG}.

\section{Message Passing Derivations}

This paper uses EP to estimate channel coefficients $\bmh_{lk}$ and BP to estimate the data symbols $x_{kt}$. Furthermore, we specify the projection family of EP in this paper to be Gaussian distributions with diagonal covariance matrices.
Now, we will examine each factor and derive its outbound message. 

The message from $\Psi_{1, kt}$ to $x_{kt}$  can be computed directly since no projection is needed, i.e., $\mu_{\Psi_{1, kt}; x_{kt}}(x_{kt})=p(x_{kt})$.

\subsection{Message from $\Psi_{2, lt}$ to $x_{kt}$ }
Following \eqref{eq:ISIT246}-\eqref{eq:EPexplainationBelief}, the extrinsic at node $\Psi_{2, lt}$ is updated by
\vspace{-2mm}
\begin{align}
&\mu_{x_{kt}; \Psi_{2, lt}}(x_{kt})\propto p(x_{kt})\prod_{l'\neq l}\mu_{\Psi_{2, l't}; x_{kt}}(x_{kt})\nonumber\\
\vspace{-2mm}
&\mu_{\bmh_{lk}; \Psi_{2, lt}}(\bmh_{lk})\propto \mu_{\Psi_{3, lg};\bmh_{lk}}(\bmh_{lk})\prod_{t'\neq t} \mu_{\Psi_{2, lt'};\bmh_{lk}}(\bmh_{lk}),
\vspace{-2mm}
\label{eq:ISIT249}
\end{align}
where the extrinsic of $\bmh_{lk}$ can be computed as a Gaussian $\mu_{\bmh_{lk}; \Psi_{2, lt}}(\bmh_{lk})=\mathcal{CN}(\bmh_{lk}|\bmm_{\bmh_{lk}; \Psi_{2, lt}}, \bmC_{\bmh_{lk}; \Psi_{2, lt}})$ with 
\vspace{-2mm}
\beq
\vspace{-2mm}
\begin{split}
    \bmC_{\bmh_{lk}; \Psi_{2, lt}}&=\left(\bmC_{\Psi_{3, lg};\bmh_{lk}}^{-1}+\sum_{t'\neq t} \bmC_{\Psi_{2, lt'};\bmh_{lk}}^{-1}\right)^{-1} \nonumber\\
    \bmm_{\bmh_{lk}; \Psi_{2, lt}}&=\bmC_{\bmh_{lk}; \Psi_{2, lt}}\left( \bmC_{\Psi_{3, lg};\bmh_{lk}}^{-1}\bmm_{\Psi_{3, lg};\bmh_{lk}} \right.\nonumber\\
    &\left.+ \sum_{t'\neq t} \bmC_{\Psi_{2, lt'};\bmh_{lk}}^{-1} \bmm_{\Psi_{2, lt'};\bmh_{lk}} \right)\nonumber
\end{split}
\eeq
According to the EP rule, the message from $\Psi_{2, lt}$ to $x_{kt}$ is
\vspace{-2mm}
\beq
\vspace{-2mm}
    \mu_{\Psi_{2, lt}; x_{kt}}(x_{kt}) \propto \frac{\text{proj}[b_{\Psi_{2, lt}; x_{kt}}(x_{kt})]}{\mu_{x_{kt}; \Psi_{2, lt}}(x_{kt})},
\eeq
where the belief (approximated posterior) at $\Psi_{2, lt}$ is defined as $b_{\Psi_{2, lt}; x_{kt}}(x_{kt})$ with
\vspace{-2mm}
\begin{align}
    &b_{\Psi_{2, lt}; x_{kt}}(x_{kt}) 
    \!\!\propto\!\!\mu_{x_{kt}; \Psi_{2, lt}} \!(x_{kt})\! \sum_{\bmx_{\kbar t}}\!\!\int\!\! p(\bmy_{lt}|x_{kt}\bmh_{lk}+\sum_{i\neq k} x_{it}\bmh_{li}) \nonumber\\
    & \cdot \mu_{\bmh_{lk}; \Psi_{2, lt}} (\bmh_{lk})
     \prod_{i\neq k}\mu_{\bmh_{li}; \Psi_{2, lt}} (\bmh_{li}) \mu_{x_{it}; \Psi_{2, lt}} (x_{it}) d\bmH_{l}. \label{eq:isit247}
\end{align}
We use the notation $\bmx_{\kbar t}$ to denote all the elements in $\bmx_{:t}$ except the $k$-th element.
The integral (and summation) in \eqref{eq:isit247} can be considered as a marginalization operation. Furthermore, we can view the extrinsic messages as hypothetical priors. Due to CLT, we approximate $\sum_{i\neq k} x_{it}\bmh_{li}$  to a Gaussian where $x_{it}\sim\mu_{x_{it}; \Psi_{2, lt}} (x_{it})$, $\bmh_{li}\sim\mu_{\bmh_{li}; \Psi_{2, lt}} (\bmh_{li})$ \cite{zou2018concise}. 
Therefore, \eqref{eq:isit247} becomes  
\vspace{-2mm}
\begin{align}
    &b_{\Psi_{2, lt}; x_{kt}}(x_{kt})\propto \mu_{x_{kt}; \Psi_{2, lt}} (x_{kt}) \label{eq:isit248}\\
    &\cdot\!\! \int \!\!\! \int\!\! p(\bmy_{lt}|x_{kt}\bmh_{lk}+\bmz_{lkt}) \mu_{\bmz_{lkt}}(\bmz_{lkt})d\bmz_{lkt}
    \!\cdot\!\mu_{\bmh_{lk}; \Psi_{2, lt}} (\bmh_{lk})d\bmh_{lk}, \nonumber
\end{align}
where $\mu_{\bmz_{lkt}}(\bmz_{lkt})=\mathcal{CN}(\bmz_{lkt}|\bmm_{\bmz_{lkt}}, \bmC_{\bmz_{lkt}})$ with
\vspace{-1mm}
    \begin{align}
         &\bmm_{\bmz_{lkt}}=\sum_{i\neq k} m_{x_{it}; \Psi_{2, lt}}\bmm_{\bmh_{li}; \Psi_{2, lt}} \label{eq:isit2413p}\\
         \vspace{-2mm}
         &\bmC_{\bmz_{lkt}}=\sum_{i\neq k} r_{x_{it}; \Psi_{2, lt}}\bmC_{\bmh_{li}; \Psi_{2, lt}}
         \!+\! \tau_{x_{it}; \Psi_{2, lt}}\bmm_{\bmh_{it}; \Psi_{2, lt}}\bmm_{\bmh_{it}; \Psi_{2, lt}}^{H} \nonumber
         \vspace{-2mm}
    \end{align}
where $m_{x_{it}; \Psi_{2, lt}}$, $\tau_{x_{it}; \Psi_{2, lt}}$ and $r_{x_{it}; \Psi_{2, lt}}$ are the mean, variance 
 and second-order moment of the normalized message $\mu_{x_{it}; \Psi_{2, lt}}$. 
By applying the Gaussian reproduction lemma \cite{zou2018concise} and the fact that Gaussian distribution integrates to one, the belief \eqref{eq:isit248} becomes
\vspace{-2mm}
\beq
\begin{split}
    &b_{\Psi_{2, lt}; x_{kt}}(x_{kt})\propto \mathcal{CN}(\bmzero| \bmy_{lt} \!-\! \bmm_{\bmz_{lkt}} \!\!-\! x_{kt}\bmm_{\bmh_{lk};\Psi_{2, lt}}, \\
    &\bmC_{\bmv} \!+\! \bmC_{\bmz_{lkt}}\!\!+\!|x_{kt}|^2\bmC_{\bmh_{lk}; \Psi_{2, lt}})\cdot\mu_{x_{kt};\Psi_{2, lt}(x_{kt})}.
    \label{eq:isit2410}
\end{split}
\eeq
Therefore, by BP rules, the outbound message is
\vspace{-2mm}
\beq
\begin{split}
    &\mu_{\Psi_{2, lt}; x_{kt}}(x_{kt})\propto \mathcal{CN}(\bmzero| \bmy_{lt} \!-\! \bmm_{\bmz_{lkt}} \!\!-\! x_{kt}\bmm_{\bmh_{lk};\Psi_{2, lt}}, \\
    &\bmC_{\bmv} \!+\! \bmC_{\bmz_{lkt}}\!\!+\!|x_{kt}|^2\bmC_{\bmh_{lk}; \Psi_{2, lt}})
\end{split}
\label{eq:isit2411}
\eeq

\subsection{Message from $\Psi_{2, lt}$ to $\bmh_{lk}$  \label{sec:isitsubc}}
Based on EP rules \eqref{eq:ISIT246},  the message to $\bmh_{lk}$ is
\beq
    \mu_{\Psi_{2, lt}; \bmh_{lk}}(\bmh_{lk}) \propto \frac{\mathrm{proj}[b_{\Psi_{2, lt}; \bmh_{lk}}(\bmh_{lk})]}{\mu_{\bmh_{lk}; \Psi_{2, lt}}(\bmh_{lk})},
\eeq
where the belief is defined as
\vspace{-2mm}
\beq
\vspace{-2mm}
\begin{split}
    &b_{\Psi_{2, lt}; \bmh_{lk}}(\bmh_{lk})\propto\sum_{\bmx_{:t}}\int p(\bmy_{lt}|\sum_{i} x_{it}\bmh_{li}) \\
    \vspace{-2mm}
    &\cdot \prod_{i}\mu_{\bmh_{li}; \Psi_{2, lt}} (\bmh_{li})\mu_{x_{it}; \Psi_{2, lt}} (x_{it}) d\bmh_{l\kbar}
\end{split}
\label{eq:isit2412}
\eeq
We use $\bmh_{l\kbar}$ to denote all the column vectors in $\bmH_{l}$ except the $k$-th column.
By using the same approach from \eqref{eq:isit247} to \eqref{eq:isit2410}, and separating the terms that contains only $x_{kt}$ \cite{zou2018concise} \cite{karataev2024bilinear}, the belief \eqref{eq:isit2412} becomes
\vspace{-2mm}
\beq
\vspace{-2mm}
\begin{split}
    &b_{\Psi_{2, lt}; \bmh_{lk}}(\bmh_{lk})\\
    =&\E_{b_{\Psi_{2, lt};x_{kt}}}\{\mathcal{CN}[\bmh_{lk}| \bmm_{\bmhh_{lk}|x_{kt}}(x_{kt}), \bmC_{\bmhh_{lk}|x_{kt}}(x_{kt})]\}
\end{split}
\label{eq:isit2413}
\eeq
where $\bmm_{\bmhh_{lk}|x_{kt}}(\cdot)$ and $\bmC_{\bmhh_{lk}|x_{kt}}(\cdot)$ are defined as
\vspace{-1mm}
\beq
\vspace{-1mm}
\begin{split}
    &\bmC_{\bmhh_{lk}|x_{kt}}(x)=[|x|^2(\bmC_{\bmv}+\bmC_{\bmz_{lkt}})^{-1}+\bmC_{\bmh_{lk}; \Psi_{2, lt}}^{-1}]^{-1}\\
    &\bmm_{\bmhh_{lk}|x_{kt}}(x)=\bmC_{\bmhh_{lk}|x_{kt}}(x)\left[\bmC_{h_{lk};\Psi_{2, lt}}^{-1}\bmm_{h_{lk}; \Psi_{2, lt}}\right.\\
    &\left.+   |x|^2(\bmC_{\bmv}+\bmC_{\bmz_{lkt}})^{-1}\frac{\bmy_{lt}-\bmm_{\bmz_{lkt}}}{x}\right],
    \label{eq:isit2414}
\end{split}
\eeq
where $\bmC_v=\sigma_v^2\bmI$. The mean $\bmm_{\bmhh_{lk}^{2}}$ and covariance 
$\bmC_{\bmhh_{lk}^{2}}$ of the belief distribution \eqref{eq:isit2413} are
\beq
\vspace{-1mm}
\begin{split}
    &\bmm_{\bmhh_{lk}^{2}}=\E_{b_{\Psi_{2, lt};x_{kt}}}[\bmm_{\bmhh_{lk}|x_{kt}}(x_{kt})]\\
    &\bmC'_{\bmhh_{lk}^{2}}=\E_{b_{\Psi_{2, lt};x_{kt}}}[\bmC_{\bmhh_{lk}|x_{kt}}(x_{kt})\\
    &+\bmm_{\bmhh_{lk}|x_{kt}}(x_{kt})\bmm_{\bmhh_{lk}|x_{kt}}(x_{kt})^H]- \bmm_{\bmhh_{lk}^{2}}\bmm_{\bmhh_{lk}^{2}}^H.
\end{split}
\eeq
We project the belief at $\Psi_{2, lt}$ to a Gaussian with diagonal covariance matrix $\mathrm{proj}[b_{\Psi_{2, lt}; \bmh_{lk}}(\bmh_{lk})]=\mathcal{CN}(\bmh_{lk}| \bmm_{\bmhh_{lk}^2}, \bmC_{\bmhh_{lk}^{2}})$, where $\bmC_{\bmhh_{lk}^{2}}$ is a digonal matrix with the same diagonal elements as $\bmC_{\bmhh_{lk}^{2}}'$.
Finally, the message from $\Psi_{2, lt}$ to $\bmh_{lk}$ is
\vspace{-1mm}
\beq
\vspace{-1mm}
\begin{split}
    \mu_{\Psi_{2, lt}; \bmh_{lk}}(\bmh_{lk})=\mathcal{CN}(\bmh_{lk}|\bmm_{\Psi_{2, lt}; \bmh_{lk}}, \bmC_{\Psi_{2, lt}; \bmh_{lk}})\\
    \propto\frac{\mathcal{CN}(\bmh_{lk}| \bmm_{\bmhh_{lk}^2}, \bmC_{\bmhh_{lk}^{2}})}{\mathcal{CN}(\bmh_{lk}|\bmm_{h_{lk}; \Psi_{2, lt}}, \bmC_{h_{lk}; \Psi_{2, lt}})}.
\end{split}
\label{eq:ISIT2421}
\eeq

\subsection{Message form $\Psi_{3, lg}$ to $\bmh_{lk}$  \label{sec:isit24subD}}
We assume $k\in G_g$.
The extrinsic at $\Psi_{3, lg}$ is updated by 
\vspace{-2mm}
\beq
\vspace{-2mm}
    \mu_{\bmh_{lk}; \Psi_{3, lg}}(\bmh_{lk})\propto\prod_t \mu_{\Psi_{2, lt};\bmh_{lk}}(\bmh_{lk}).
    \label{eq:ISIT2420}
\eeq
We denote this extrinsic message as $\mu_{\bmh_{lk}; \Psi_{3, lg}}(\bmh_{lk})=\mathcal{CN}(\bmh_{lk}|\bmm_{\bmh_{lk}; \Psi_{3, lg}}, \bmC_{\bmh_{lk}; \Psi_{3, lg}})$ with
\vspace{-2mm}
\beq
\vspace{-1mm}
\begin{split}
    \bmC_{\bmh_{lk}; \Psi_{3, lg}}&=\left(\sum_{t} \bmC_{\Psi_{2, lt};\bmh_{lk}}^{-1}\right)^{-1} \nonumber\\
    \bmm_{\bmh_{lk}; \Psi_{3, lg}}&=\bmC_{\bmh_{lk}; \Psi_{3, lg}}\left(
    \sum_{t} \bmC_{\Psi_{2, lt};\bmh_{lk}}^{-1} \bmm_{\Psi_{2, lt};\bmh_{lk}} \right).\nonumber
\end{split}
\eeq
The belief of $\bmh_{lg}$ at the $\Psi_{3, lg}$ is
\beq
    b_{\Psi_{3, lg}}(\bmh_{lg})\propto p(\bmy_{p, lg}, \bmh_{lg}) \prod_{k\in G_g}p(\bmh_{lk})\mu_{\bmh_{lk}; \Psi_{3, lg}}(\bmh_{lk}).
    \label{eq:spawc2429}
\eeq
All the factors appearing in \eqref{eq:spawc2429} are Gaussian pdfs with diagonal covariance matrices. Therefore, the projection of $b_{\Psi_{3, lg}}(\bmh_{lk})$ results to itself. For simplicity, we define a hypothetical prior $q_{\bmh_{lk}|\bmY_d}$  for $\bmh_{lk}$ in \eqref{eq:spawc2429} as
\vspace{-2mm}
\beq
\vspace{-1mm}
\begin{split}
    q_{\bmh_{lk}|\bmY_d}(\bmh_{lk})=\mathcal{N}(\bmh_{lk}|\bmm_{\bmh_{lk}|\bmY_d}, \bmC_{\bmh_{lk}|\bmY_d})\\
    \propto p(\bmh_{lk})\mu_{\bmh_{lk}; \Psi_{3, lg}}(\bmh_{lk}),
\end{split}
\eeq
where
\vspace{-2mm}
\beq
\vspace{-1mm}
\begin{split}
    \bmC_{\bmh_{lk}|\bmY_d}=(\bmXi_{\bmh_{lk}}^{-1}+\bmC_{\bmh_{lk};\Psi_{3, lg}}^{-1})^{-1}\\
    \bmm_{\bmh_{lk}|\bmY_d}=\bmC_{\bmh_{lk}|\bmY_d}\bmC_{\bmh_{lk};\Psi_{3, lg}}^{-1}\bmm_{\bmh_{lk};\Psi_{3, lg}}
\end{split}
\label{eq:ISIT2424}
\eeq
The message from factor node $\Psi_{3, lg}$ to $\bmh_{lk}$ can be derived as
\vspace{-1.5mm}
\beq
\vspace{-1mm}
\begin{split}
    &\mu_{\Psi_{3, lg}; \bmh_{lk}}(\bmh_{lk})\propto \frac{\int b_{\Psi_{3, lg}}(\bmh_{lg}) d \bmh_{l\kbar}}{\mu_{\bmh_{lk}; \Psi_{3, lg}}(\bmh_{lk})}\\
    &\propto p(\bmh_{lk})\frac{\int p(\bmy_{p, lg}, \bmh_{lg}) \prod_{k\in G_g}q_{\bmh_{lk}|\bmY_d}(\bmh_{lk}) d \bmh_{l\kbar}}{q_{\bmh_{lk}|\bmY_d}(\bmh_{lk})}
\end{split}
\label{eq:ISIT2425}
\eeq
The fraction operation in the second line of \eqref{eq:ISIT2425} can be interpreted as component-wise conditionally-unbiased LMMSE estimation \cite{EURECOM+7412}. Therefore, the message from $\Psi_{3, lg}$ to $\bmh_{lk}$ is
\vspace{-1mm}
\beq
\vspace{-0mm}
    \mu_{\Psi_{3, lg}; \bmh_{lk}}(\bmh_{lk})=\mathcal{CN}(\bmh_{lk}|\bmm_{\Psi_{3, lg}; \bmh_{lk}}, \bmC_{\Psi_{3, lg}; \bmh_{lk}}),
    \label{eq:isit2425}
\eeq
where
\vspace{-4mm}
\begin{align}
\vspace{-2mm}
    &\bmC_{\Psi_{3, lg}; \bmh_{lk}}\!=\!\left[\bmXi_{\bmh_{lk}}^{-1}+\left(\frac{\sigma_v^2}{\sigma_x^2P}\bmI+\sum_{k'\in G_g/\{k\}}\bmC_{\bmh_{lk'}|\bmY_d}\right)^{-1}\right]^{-1} \nonumber\\
    &\bmm_{\Psi_{3, lg};\bmh_{lk}}\!=\!\bmXi_{\bmh_{lk}}\left(\frac{\sigma_v^2}{\sigma_x^2P}\bmI+\sum_{k'\in G_g/\{k\}}\bmC_{\bmh_{lk'}|\bmY_d}+\bmXi_{\bmh_{lk}}\right)^{-1}\nonumber\\
    &\cdot\left(\frac{1}{\sigma_x^2P}\bmy_{p, lg}-\sum_{k'\in G_g/\{k\}}\bmm_{\bmh_{lk'}|\bmY_d}\right).\label{eq:spawc24133}
\end{align}

\section {Asymptotic Behaviors in Large Systems}
For scalable systems, $L\to \infty$ while $K=c_1 L$, $P=c_2 L$, $T=c_3 L$,  where $c_1$, $c_2$, $c_3$ are some positive constants, we assume the channel coefficients $\forall l,n, k, \E[|h_{lnk}|^2]=O(\frac{1}{L})$ data constellation symbols $\forall s\in\mathcal{S}, s=O(1), 1/s=O(1)$. Furthermore, we assume the noise power scales as $\sigma_v^2=O(1)$, $\sigma_v^{-2}=O(1)$. For simplicity, we define big-O-notations with matrix parameters to represent the element-wise asymptotic behavior, i.e.,  for matrices $\bmA$, $\bmB$ of the same size, we have $\bmA=O(\bmB) \Leftrightarrow \forall i, j, [\bmA]_{ij}=O([\bmB]_{ij})$.

\begin{assumption}
\label{th:assumption1}
We assume that $\E(\bmm_{\bmh_{lk}; \bmPsi_{2, lt}})=\bmzero$, $\E[m_{x_{it}; \Psi_{2, lt}}\bmm_{\bmh_{li}; \Psi_{2, lt}}]=\bmzero$, and $\forall i\neq j, \E([\bmm_{\bmh_{lk}; \bmPsi_{2, lt}}\bmm_{\bmh_{lk}; \bmPsi_{2, lt}}^H]_{ij})=0$.
\end{assumption}
\begin{property}
    For invertible matrices $\bmA, \bmB$, we have $(\bmA^{-1}+\bmB^{-1})^{-1}=\bmA(\bmA+\bmB)^{-1}\bmB=\bmB(\bmA+\bmB)^{-1}\bmA$.
\label{pp:isit241}
\end{property}

\begin{lemma}
With proper initialization and Assumption \ref{th:assumption1}, in each iteration, the updates satisfy $\bmm_{\bmz_, lkt}=O(\bmone)$, $\bmC_{\bmz_{lkt}}=O(\bmI)+O(\frac{\bmone\cdot\bmone^H}{\sqrt L})$, $\bmm_{\bmhh_{lk}|x_{kt}}(x)=O(\frac{\bmone}{\sqrt L})$, $\bmC_{\bmhh_{lk}|x_{kt}}(x)=O(\frac{\bmI}{L})+O(\frac{\bmone\cdot\bmone^H}{L^2})$, $\bmm_{\Psi_{2, lt}; \bmh_{lk}}=O(\frac{\bmone}{\sqrt{L}})$, $\bmC_{\Psi_{2, lt}; \bmh_{lk}}=O(\bmI)$, $\bmm_{\Psi_{3, lg};\bmh_{lk}}=O(\frac{\bmone}{\sqrt{L}})$, $\bmC_{\Psi_{3, lg}; \bmh_{lk}}=O(\frac{\bmI}{L})$, $\bmm_{\bmh_{lk}; \Psi_{2, lt}}=O(\frac{\bmone}{\sqrt L})$, $\bmC_{\bmh_{lk}; \Psi_{2, lt}}=O(\frac{\bmI}{L})$. Furthermore, $\bmC_{\Psi_{2, lt}; \bmh_{lk}}^{-1}=O(\bmI)$.
\end{lemma}
\begin{proof}
    We prove this lemma by mathematical induction. Due to a proper initialization, we can assume the messages $\bmm_{\bmh_{lk}; \Psi_{2, lt}}$, $\bmC_{\bmh_{lk}; \Psi_{2, lt}}$, $\bmm_{\Psi_{2, lt}; \bmh_{lk}}$, $\bmC_{\Psi_{2, lt}; \bmh_{lk}}$, $\bmm_{\Psi_{3, lg};\bmh_{lk}}$, $\bmC_{\Psi_{3, lg}; \bmh_{lk}}$ are initialized with the above-mentioned scales.
    
    Then, we assume the lemma holds for the previous iterations and investigate the updates in the next iteration.

    We first look at the update of $\bmm_{\bmz_, lkt}$, $\bmC_{\bmz_{lkt}}$, which are updated according to \eqref{eq:isit2413p}. Similar to \cite{schniter2020simple}, we assume that $\forall i, m_{x_{it};\Psi_{2, lt}}$ are weakly independent of $\bmm_{\bmh_{li}; \Psi_{2, lt}}$. Since the elements in the constellation set scale with $O(1)$, we know $m_{x_{it}; \Psi_{2, lt}}=O(1)$. According to induction assumptions, the extrinsic mean $\bmm_{\bmh_{li};\Psi_{2, lt}}=O(\frac{\bmone}{\sqrt{L}})$. Due to Assumption \ref{th:assumption1}, we use the results from \cite[Lemma 1]{schniter2020simple} to obtain $\bmm_{\bmz_{lkt}}=\sum_{i\neq k} m_{x_{it}; \Psi_{2, lt}}\bmm_{\bmh_{li}; \Psi_{2, lt}}=O(\bmone)$ and the covariance matrix $\bmC_{\bmz_{lkt}}= O({\bmI}) + O(\frac{\bmone\cdot\bmone^H}{\sqrt L})$. For simplicity, we denote the diagonal terms of $\bmC_{\bmz_{lkt}}$ in \eqref{eq:isit2413p} as $\bmD_{\bmz_{lkt}}=\sum_{i\neq k} r_{x_{it}; \Psi_{2, lt}}\bmC_{\bmh_{li}; \Psi_{2, lt}}$, and denote $\bmB_{\bmz_{lkt}}=\sum_{i\neq k}\tau_{x_{it}; \Psi_{2, lt}}\bmm_{\bmh_{it}; \Psi_{2, lt}}\bmm_{\bmh_{it}; \Psi_{2, lt}}^{H}=O({\bmI}) + O(\frac{\bmone\cdot\bmone^H}{\sqrt L})$. Thus, with these notations, $\bmC_{\bmz_{lkt}}=\bmD_{\bmz_{lkt}}+\bmB_{\bmz_{lkt}}$.

    Now we investigate the update of $\bmm_{\bmhh_{lk}|x_{kt}}(x)$, $\bmC_{\bmhh_{lk}|x_{kt}}(x)$ in \eqref{eq:isit2414}. By matrix inversion lemma,
    \vspace{-2mm}
    \beq
    \vspace{-2mm}
        \begin{split}
            &\bmC_{\bmhh_{lk}|x_{kt}}(x)=\bmC_{\bmh_{lk}; \Psi_{2, lt}}\\
            &-\!\bmC_{\bmh_{lk}; \Psi_{2, lt}}\bmQ_{lkt}^{-\frac{H}{2}}\bmW_{lkt}\!\left(\bmLambda_{lkt}\!+\!\bmI\right)^{-1}\!\bmW_{lkt}^H\bmQ_{lkt}^{-\frac{1}{2}}\bmC_{\bmh_{lk}; \Psi_{2, lt}},
        \end{split}
        \label{eq:isit2427}
    \eeq
    where we define the positive semi-definite diagonal matrix $\bmQ_{lkt}=\bmC_{\bmh_{lk}; \Psi_{2, lt}}+\frac{1}{|x|^2}(\bmC_\bmv+\bmD_{\bmz_{lkt}})=O(\bmI)$ and $\bmQ_{lkt}^{\frac{1}{2}}\bmQ_{lkt}^{\frac{H}{2}}=\bmQ_{lkt}$.
    By eigendecomposition, we define $\frac{1}{|x|^2}\bmQ_{lkt}^{-\frac{1}{2}}\bmB_{\bmz_{lkt}}\bmQ_{lkt}^{-\frac{H}{2}}=\bmW_{lkt}\bmLambda_{lkt}\bmW_{lkt}^H$, $\bmI=\bmW_{lkt}\bmW_{lkt}^H$. Therefore, $\bmC_{\bmhh_{lk}|x_{kt}}(x)=O(\frac{\bmI}{L})+O(\frac{\bmone\cdot\bmone^H}{L^2})$. We find the update of $\bmm_{\bmhh_{lk}|x_{kt}}$ in \eqref{eq:isit2414} is dominated by the first term $\bmC_{\bmhh_{lk}|x_{kt}}(x)\bmC_{\bmh_{lk};\Psi_{2, lt}}^{-1}\bmm_{\bmh_{lk}; \Psi_{2, lt}}$. By neglecting higher order infinitesimal terms, we have
    \vspace{-3mm}
    \beq
    \vspace{-2mm}
    \begin{split}
    &\bmm_{\bmhh_{lk}|x_{kt}}\simeq \bmC_{\bmhh_{lk}|x_{kt}}(x)\bmC_{h_{lk};\Psi_{2, lt}}^{-1}\bmm_{h_{lk}; \Psi_{2, lt}}
    =[\bmI
    +\bmC_{\bmh_{lk};\Psi_{2, lt}}^{\frac{1}{2}}\\
    &\cdot(\bmC_{\bmh_{lk};\Psi_{2, lt}}^{\frac{H}{2}}\!\!\bmC_{\text{in}, lkt}^{-1}\bmC_{\bmh_{lk};\Psi_{2, lt}}^{\frac{1}{2}}\!\!\!+\!\bmI)^{-1}\bmC_{\bmh_{lk};\Psi_{2, lt}}^{-\frac{H}{2}}\bmC_{\text{in}, lkt}^{-1}]\bmm_{\bmh_{lk};\Psi_{2, lt}}\\
    &\simeq \bmm_{\bmh_{lk}; \Psi_{2, lt}}=O(\frac{\bmone}{\sqrt{L}}). \nonumber
    \end{split}
    \eeq
    To study the messages $\bmm_{\Psi_{2, lt}; \bmh_{lk}}$, $\bmC_{\Psi_{2, lt}; \bmh_{lk}}$, we first investigate the approximated (projected) belief of $\bmh_{lk}$ at $\Psi_{2, lt}$. From the previous discussion,  
    $\bmC'_{\bmhh_{lk}^{2}}\simeq \E_{b_{\Psi_{2, lt};x_{kt}}}[\bmC_{\bmhh_{lk}|x_{kt}}(x_{kt})]$, and thus,
    \vspace{-1mm}
    \beq
    \vspace{-2.5mm}
    \begin{split}
    &\bmC'_{\bmhh_{lk}^{2}}\simeq \bmC_{\bmh_{lk}; \Psi_{2, lt}}-\bmC_{\bmh_{lk}; \Psi_{2, lt}} 
    \cdot \bmF\cdot\bmC_{\bmh_{lk}; \Psi_{2, lt}},
    \end{split}
    \eeq
    where $
    \bmF=\sum_{x\in \mathcal{S}}b_{\Psi_{2, lt}; x_{kt}}(x)\left[\frac{1}{|x|^2}\bmB_{\bmz_{lkt}}
    +\bmQ_{lkt}\right]^{-1}.
    $
    Thanks to the projection in EP, we are only interested in the diagonal elements of $\bmC'_{\bmhh_{lk}^{2}}$. With the approximation $\bmm_{\bmhh_{lk}|x_{kt}}\simeq \bmm_{\bmh_{lk};\Psi_{2, lt}}$, the $n$-th diagonal term reads
    \vspace{-2mm}
    \begin{align}
        &[\bmC'_{\bmhh_{lk}^{2}}]_{nn}=[\bmC_{\bmhh_{lk}^{2}}]_{nn}\simeq\tau_{h_{lnk};\Psi_{2, lt}}-\tau_{h_{lnk};\Psi_{2, lt}}^2[\bmF]_{nn}\nonumber\\
        &=[([\bmF]_{nn}^{-1}-\tau_{h_{lnk};\Psi_{2, lt}})^{-1}+\tau_{h_{lnk};\Psi_{2, lt}}^{-1}]^{-1}
        \label{eq:ISIT2431}
    \end{align}
    From the same analysis in \eqref{eq:isit2427}, we know $\bmQ_{lkt}$ is asymptotic upper and lower bounded. Since $\bmB_{\bmz_{lkt}}$ is positive semi-definite, we have $[\bmF]_{nn}=O(1)$ and $[\bmF]_{nn}^{-1}=O(1)$. Substitute \eqref{eq:ISIT2431} into \eqref{eq:ISIT2421}, and we obtain $[\bmC_{\Psi_{2, lt}; \bmh_{lk}}]_{nn}\simeq [\bmF]_{nn}^{-1}-\tau_{h_{lnk};\Psi_{2, lt}}$. Thus, $\bmC_{\Psi_{2, lt}; \bmh_{lk}}=O(\bmI)$ and $\bmC_{\Psi_{2, lt}; \bmh_{lk}}^{-1}=O(\bmI)$. Since $\bmm_{\bmhh_{lk}|x_{kt}}\simeq \bmm_{h_{lk}; \Psi_{2, lt}}$, it is straightforward to see $\bmm_{\Psi_{2, lt};\bmh_{lk}}=O(\frac{\bmone}{\sqrt L})$.

    The message covariance matrices $\bmC_{\Psi_{3, lg};\bmh_{lk}}$ and $\bmC_{\bmh_{lk};\Psi_{2, lt}}$ are both diagonal matrices. Due to the scalable-system assumption $T=O(L)$ and the elements in $\bmC_{\Psi_{2, lt};\bmh_{lk}}$ being asymptotically upper and lower bounded, one can show $\bmC_{\Psi_{3, lg};\bmh_{lk}}=O(\frac{\bmI}{L})$ is upper bounded, and $\bmm_{\Psi_{3, lg}; \bmh_{lk}}=O(\frac{\bmone}{\sqrt{L}})$ according to \eqref{eq:ISIT2424}-\eqref{eq:spawc24133}. We can then show $\bmC_{\bmh_{lk};\Psi_{2, lt}}=O(\frac{\bmI}{L})$ and $\bmm_{\bmh_{lk};\Psi_{2, lt}}=O(\frac{\bmone}{\sqrt{L}})$ according to \eqref{eq:ISIT249}.
\end{proof}

\section {Simplication of the Messages}

We define beliefs at the variable nodes as
\vspace{-1mm}
\beq
\vspace{-2mm}
\begin{split}
\vspace{-2mm}
    b_{x_{kt}}(x_{kt})\propto p(x_{kt})\prod_{l}\mu_{\Psi_{2, lt}; x_{kt}}(x_{kt})\\
    \vspace{-2mm}
    b_{\bmh_{lk}}(\bmh_{lk})\propto \mu_{\Psi_{3, lg};\bmh_{lk}}(\bmh_{lk})\prod_t \mu_{\Psi_{2, lt};\bmh_{lk}}(\bmh_{lk}).
    \vspace{-2mm}
    \label{eq:ISIT2430}
\end{split}
\eeq
Compared to the extrinsic messages in \eqref{eq:ISIT249}, the beliefs in \eqref{eq:ISIT2430} only differ by one factor. Since Loopy BP is used for estimating $x_{kt}$, it has been shown in \cite{parker2014bilinear} that we can assume $b_{x_{kt}}(x_{kt})\simeq\mu_{x_{kt}; \Psi_{2, lt}}(x_{kt})$.

This work estimates the channel coefficients $\bmh_{lk}$ using EP. Therefore, a separate analysis from \cite{parker2014bilinear} is needed. We investigate the mean and covariance matrix difference between $b_{\bmh_{lk}}(\bmh_{lk})$ and $\mu_{\bmh_{lk};\Psi_{2, lt}}$ based on the Lemma 2.

Substitute \eqref{eq:ISIT249} into \eqref{eq:ISIT2430}, and we obtain the belief at $\bmh_{lk}$ as
\vspace{-2mm}
\beq
\vspace{-2mm}
    b_{\bmh_{lk}}(\bmh_{lk})\propto \mu_{\bmh_{lk}; \Psi_{2, lt}}(\bmh_{lk})\mu_{\Psi_{2, lt};\bmh_{lk}}(\bmh_{lk}).
\eeq
Denote $\bmm_{\bmhh_{lk}}$ and $\bmC_{\bmhh_{lk}}$ as the mean and covariance matrix of $b_{\bmh_{lk}}(\bmh_{lk})$. From Lemma 2, we have 
\vspace{-1mm}
\begin{align}
    &\bmC_{\bmh_{lk};\Psi_{2, lt}} \!\!\!-\! \bmC_{\bmhh_{lk}} \!\!\!\!=\! \bmC_{\bmh_{lk};\Psi_{2, lt}}^2\!(\bmC_{\bmh_{lk};\Psi_{2, lt}}\!\!\!+\!\bmC_{\Psi_{2, lt};\bmh_{lk}})^{-1} \nonumber \\
    &\bmm_{\bmh_{lk};\Psi_{2, lt}} - \bmm_{\bmhh_{lk}} = \bmC_{\bmh_{lk};\Psi_{2, lt}}(\bmC_{\bmh_{lk};\Psi_{2, lt}}+\bmC_{\Psi_{2, lt};\bmh_{lk}})^{-1} \nonumber\\
    &\cdot(\bmm_{\bmh_{lk};\Psi_{2, lt}}-\bmm_{\Psi_{2, lt}; \bmh_{lk}}).
    \label{eq:ISIT2432p}
\end{align}
It has been shown in the proof of Lemma 2 that the difference $\bmm_{\bmh_{lk};\Psi_{2, lt}}-\bmm_{\Psi_{2, lt}; \bmh_{lk}}$ is a higher order infinitesimal relative to $\bmm_{\bmh_{lk};\Psi_{2, lt}}$. Thus, the quotients $(\bmC_{\bmh_{lk};\Psi_{2, lt}} \!\!\!-\! \bmC_{\bmhh_{lk}})\bmC_{\bmh_{lk};\Psi_{2, lt}}^{-1}$ and $(\bmm_{\bmh_{lk};\Psi_{2, lt}} - \bmm_{\bmhh_{lk}})./\bmm_{\bmh_{lk};\Psi_{2, lt}}$ tend to zero as the system grows larger. Therefore, the difference in \eqref{eq:ISIT2432p} are higher order infinitesimals relative to $\bmC_{\bmh_{lk};\Psi_{2, lt}}$ and $\bmm_{\bmh_{lk};\Psi_{2, lt}}$, respectively. Therefore, we have $b_{\bmh_{lk}}(\bmh_{lk})\simeq \mu_{\bmh_{lk}; \Psi_{2, lt}}(\bmh_{lk})$. 
Based on the above discussion, we propose to replace the extrinsic \eqref{eq:ISIT249} at $\Psi_{2, lt}$ by \eqref{eq:ISIT2432} and \eqref{eq:ISIT2433} to reduce complexity further,
\vspace{-2mm}
\begin{align}
    &\mu_{\bmh_{lk}; \Psi_{2, lt}}'(\bmh_{lk})=b_{\bmh_{lk}}(\bmh_{lk}); \label{eq:ISIT2432}\\
    &\mu_{x_{kt}; \Psi_{2, lt}}'(x_{kt})=b_{x_{kt}}(x_{kt}). \label{eq:ISIT2433}
\end{align}

\section{Decentralized Method}
To obtain the belief of $x_{kt}$, we need to combine the message from all the AP.
We consider the case where all the $L$ AP are connected, and the AP network has a tree structure. A decentralized message-passing method can be used based on the consensus propagation framework \cite{moallemi2006consensus}. 
Define the normalized message from AP $l$ to AP $l'$:
\vspace{-2mm}
\beq
\vspace{-2mm}
    \nu_{l\to l'} (x_{kt})\propto \mu_{\Psi_{2, lt}; x_{kt}}(x_{kt})\prod_{\lpbar \in N(l)/ \{l'\} }\nu_{\lpbar\to l}(x_{kt}),\nonumber
\eeq
where $N(l)$ denotes the set of connected neighbors of AP $l$.
At convergence, the belief in \eqref{eq:ISIT2430} can be obtained by any AP $l$ as
\vspace{-1mm}
\beq
\vspace{-1mm}
    b'_{x_{kt}}(x_{kt})\propto p(x_{kt})\mu_{\Psi_{2, lt}; x_{kt}}(x_{kt})\prod_{l' \in N(l)}\nu_{l'\to l}(x_{kt}).
    \label{eq:ISIT2436}
\eeq
Therefore, for a decentralized algorithm, we can replace the update of belief $b_{x_{kt}}$ in \eqref{eq:ISIT2430} by $b'_{x_{kt}}$ in \eqref{eq:ISIT2436}.
After updating the message $\mu_{\Psi_{2, lt}; x_{kt}}$, we update the shared message by
\vspace{-1mm}
\beq
\vspace{-1mm}
    \nu_{l\to l'}^{new} (x_{kt})\propto \mu_{\Psi_{2, lt}; x_{kt}}^{new}(x_{kt})\prod_{\lpbar \in N(l)/ \{l'\} }\nu_{\lpbar\to l}^{old}(x_{kt}),
    \label{eq:ISIT2436p}
\eeq
where we use $new$ and $old$ to distinguish the message of different iterations.
One possible ordering method is suggested in Algorithm \ref{algo:isit241}. 

\floatstyle{spaceruled}
\restylefloat{algorithm}
\begin{algorithm}[t]
\caption{One Iteration of Decentralized EP}\label{algo:isit241}
\begin{algorithmic}[1]
\Require $\bmXi_{\bmh_{lk}}$, $\bmy_{p, lg}$ , $\bmy_{lt}$, $p(x_{kt})$, $\sigma_x^2$, $\sigma_v^2$, $G_g$\\
Initialize $\mu_{\Psi_{3, lg};\bmh_{lk}}$, $\mu_{\Psi_{2, lt};x_{kt}}$, $\mu_{\Psi_{2, lt};\bmh_{lk}}$, $\nu_{l\to l'}(x_{kt})$
\State At all the APs, $\forall k, t$, update $b_{x_{kt}}$ according to \eqref{eq:ISIT2436}
\For {l=1:L}
\State $\forall k$, update $\mu_{\bmh_{lk};\Psi_{3, lg}}$ based on \eqref{eq:ISIT2420}
\State $\forall k, t$, update $\mu'_{\bmh_{lk};\Psi_{2, lt}}$ based on \eqref{eq:ISIT2430} and \eqref{eq:ISIT2433}
\State $\forall k, t$, update $\mu'_{x_{kt};\Psi_{2, lt}}$ based on \eqref{eq:ISIT2432}
\State $\forall k$, update $\mu_{\Psi_{3, lg};\bmh_{lk}}$ based on \eqref{eq:isit2425}-\eqref{eq:spawc24133}
\State $\forall k, t$, update $\mu_{\Psi_{2, lt};x_{kt}}$ based on \eqref{eq:isit2413p}-\eqref{eq:isit2411}
\State $\forall k, t$, update $\mu_{\Psi_{2, lt};\bmh_{lk}}$ based on \eqref{eq:isit2414}-\eqref{eq:ISIT2421}
\State $\forall l'\in N(l), k, t$, update $\nu_{l\to l'}(x_{kt})$ based on \eqref{eq:ISIT2436p}
\EndFor
\end{algorithmic}
\end{algorithm}

\vspace{-1mm}
\section{Simulation Results}
\vspace{-1mm}
Our study simulates an environment within a $400 \times 400$ square meter area, equipped with $16$ APs and $8$ User Terminals (UTs). Each AP features $N=2$ antennas and is positioned at coordinates $(\frac{400}{3} i, \frac{400}{3} j)$, $i, j\in \{0, 1, 2, 3\}$. The UTs are uniformly distributed throughout the area. We denote the distance between each UT $k$ and AP $l$ as $d_{lk}$. Channel covariances for each user $k$ at AP $l$ are modeled using $N \times N$ diagonal matrices, represented as $\sigma_{h_{lk}}^2\bmI$, where $10\log_{10}(\sigma_{h_{lk}}^2)=-30 - 36.7\log_{10}(d_{lk})$.

All the neighboring APs within $\frac{400}{3}$ meters are connected and can exchange information of the estimated data symbols. Furthermore, as illustrated in Algorithm \ref{algo:isit241}, a synchronized message-exchanging scheme is used.

The length of the orthogonal pilot sequences is set to $P=6$ to introduce pilot contamination.

\begin{figure}[t]
    \centering
    \includegraphics[width=0.485\textwidth]{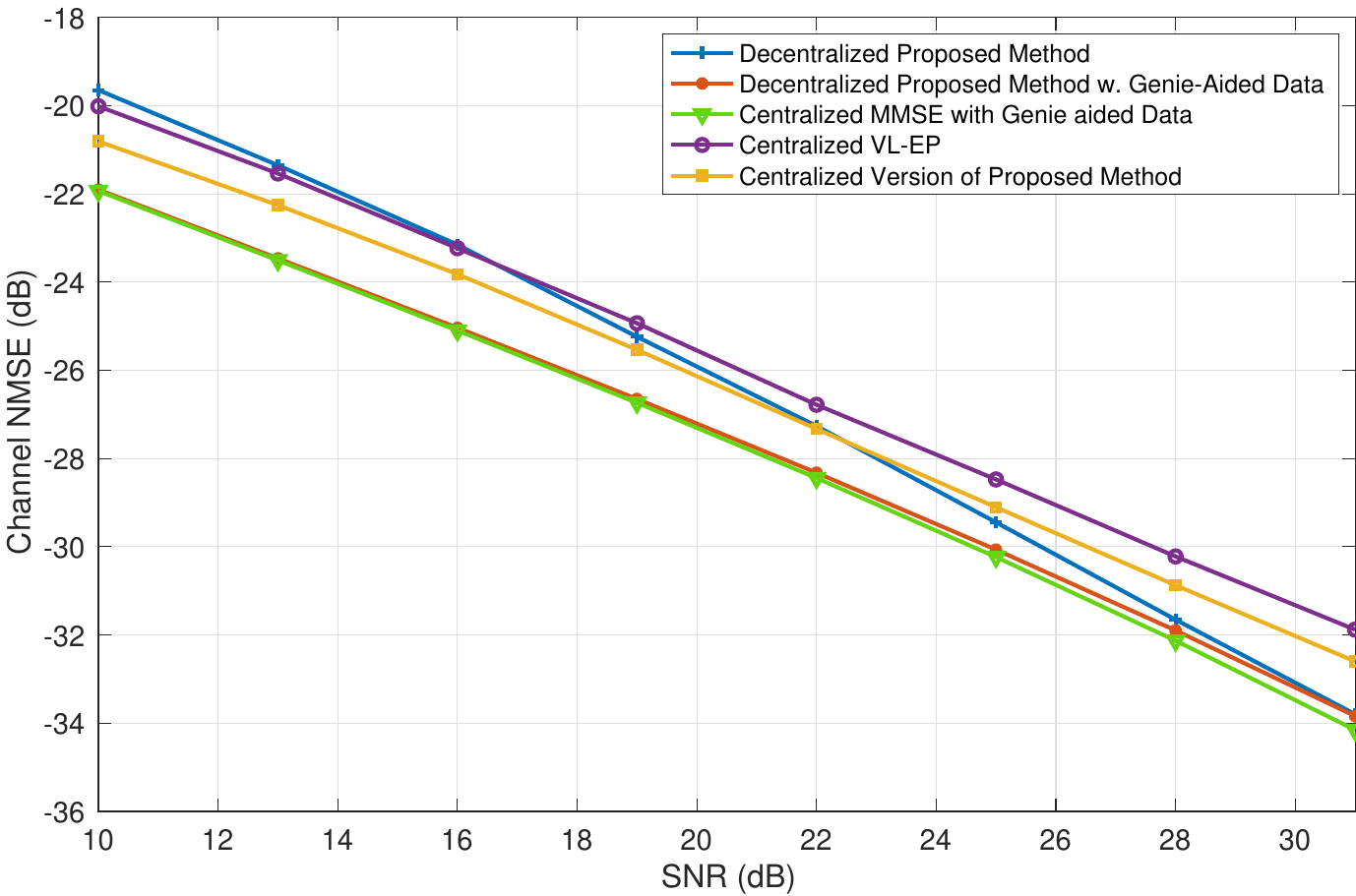}
    \vspace{-6mm}
    \caption{NMSE vs SNR} 
    \vspace{-2mm}
    \label{fig:N}
    \vspace{-4mm}
\end{figure}

We employ a $4$QAM constellation of length $T=10$ for signal transmission and assume a noise power of $-96$ dBm. The signal-to-noise ratio (SNR) is adjusted by varying the transmitted power. We base our results on $100$ different realizations, which are illustrated in Figure \ref{fig:N}. 
The normalized mean squared error (NMSE) of the channel estimates is defined as
$
    \mathrm{NMSE}=\frac{\tr[|(\bmHh-\bmH)|^2]}{\tr[|\bmH|^2]}
$,
where $\bmHh$ are synthesized from the mean of $b_{\bmh_{lk}}(\bmh_{lk})$ defined in \eqref{eq:ISIT2430} and the operation $|\cdot|^2$ is defined as $|\bmH|^2=\bmH^H\bmH$.

In the VL-EP scenario, we generate data symbols drawn from i.i.d. Gaussian distribution and apply the VL-EP algorithm \cite{gholami2021message} for channel estimation. In the Genie-Aided scenario, we implement the proposed algorithm as if the data symbols are known. In the MMSE Genie-Aided scenario, all the APs estimate the channel coefficients using the MMSE estimator with known channel coefficients.

\vspace{-1mm}
\section{Conclusions}
\vspace{-1mm}
This paper introduces a simplified, decentralized EP-based algorithm for bilinear joint estimation. To simplify the factorization scheme, we leverage orthogonal pilots and the CLT. Through asymptotic analysis, we further refine the message update scheme within the algorithm. Although originally developed for an acyclic network of APs, our simulation results confirm the algorithm's effectiveness even when the APs are interconnected in a cyclic network.

{\bf Acknowledgements}
EURECOM's research is partially supported by its industrial members:
ORANGE, BMW, SAP, iABG,  Norton LifeLock, and the Franco-German project CellFree6G.
\vspace{-4mm}

\bibliographystyle{IEEEtran}

\bibliography{isit24_ref}


\end{document}